# Multi-megawatt, self-seeded Mamyshev oscillator

PAVEL SIDORENKO*, WALTER FU, LOGAN G WRIGHT AND FRANK W WISE

*School of Applied and Engineering Physics, Cornell University, Ithaca, New York 14853, USA*
*Corresponding author: ps865@cornell.edu*

**Abstract:** We demonstrate a fiber oscillator that achieves 3 MW peak power, is easily started, and is environmentally stable. The Mamyshev oscillator delivers 190-nJ pulses that can be compressed externally to 35 fs duration. Accurate numerical modeling of the gain medium provides insight into the behavior and performance of the device.

Ultrafast fiber lasers attract considerable interest as an alternative to bulk solid-state lasers due to their efficiency, compactness, and advantages of the waveguide medium. However, the pulse energy is typically limited by the increasing difficulty of stabilizing high-energy, single-pulse evolution.

Pulse evolutions based on a normal-dispersion fiber provide a means of tolerating high nonlinear phase accumulation without sacrificing pulse quality [1]. However, excessive nonlinear phase accumulation causes a pulse in the cavity to split apart (i.e., enter a multi-pulsing regime) and/or spectral/temporal pulse shape distortions. In parallel with this issue, pulses that accumulate high nonlinear phase shifts need to be stabilized by a proportionally-strong saturable absorber (SA) [2–4]. If the SA's modulation depth is too shallow, the pulse evolution can be disrupted by the growth of continuous-wave background (termed CW breakthrough). The SA also plays the vital role of initiating pulse formation from noise.

Saturable absorbers based on material absorption [5–7] and interference-based SAs [8–10] have been employed in fiber oscillators. Unfortunately, none of these can provide a perfect solution for high-energy pulse stabilization. Material-based SAs are prone to damage in high-power fiber lasers [11,12]. The transmission-intensity curves of interference-based SAs do not increase monotonically [8–10], and lasers with these tend to emit multi-pulse bursts at the highest energies. Furthermore, only SAs based on nonlinear polarization evolution (NPE) have demonstrated deep enough modulation (often >70%) to support high-energy pulses. However, the use of low-birefringence fibers makes such lasers sensitive to environmental perturbations.

One promising solution has its roots in Mamyshev's proposal to regenerate pulses using spectral broadening followed by offset spectral filtering [13]. While the idea of a mode-locked laser based on offset filters was proposed much earlier [14,15], it did not receive significant attention at that time. More recently, several approaches for short-pulse generation based on the Mamyshev regenerator have been investigated [16–18]. This approach has gained much more momentum in the past few years, with the demonstration of mode-locked oscillators based on two concatenated Mamyshev regenerators (so-called Mamyshev oscillators) that have achieved nanojoule-scale pulse energies [19–21]. This approach permits both unprecedented performance and an environmentally-stable design. The outstanding performance of the Mamyshev oscillator relies primarily on the effective saturable absorber created by the concatenated Mamyshev regenerators. Increasing the filter separation increases the modulation depth [16,22], which allows higher-energy pulses to be stabilized. However, the resulting suppression of low-intensity fluctuations also hinders starting of the oscillator. With small filter separations (~3 nm), the low-intensity attenuation is mild enough to permit starting through signal modulation [19]. Using more moderate (~10 nm) filter separations permits higher steady-state energies, but requires stronger starting methods such as modulation of the pump power [20]. To achieve the highest pulse energies to-date, a large enough filter separation is required that none of these starting mechanisms suffice [21]. Starting that laser requires an ultrashort external seed pulse, which significantly diminishes the attractiveness of the Mamyshev oscillator as a practical device.

Here, we present a simple and reliable method for starting a Mamyshev oscillator that generates pulses with multi-megawatt peak power. The oscillator includes a sub-cavity that creates a fluctuating field that seeds the mode-locked state. Moreover, we scale the cavity design to a larger fiber core compared to previous work [21] to generate stable, 190-nJ pulses that are dechirped to 35 fs duration, yielding 3 MW peak power. With the incorporated starting cavity, the oscillator is self-seeded, and is environmentally-stable in the mode-locked state.

The oscillator design is based on the normal-dispersion ring cavity presented in [21], with polarization-maintaining (PM) Yb-doped fiber as the gain medium. The cavity (Fig. 1) is constructed by concatenating two Mamyshev regenerators (the two "arms"). The first arm (red dashed box in Fig. 1) is built from fiber with a 6-µm core diameter, and acts mainly as a lower-energy feedback loop for the second arm. The second arm (blue dashed box in Fig. 1) is constructed from fiber with a 10-µm core diameter, and acts as a power amplifier after which the main output is taken. The offset bandpass filters that complete the regenerators are placed between the arms.

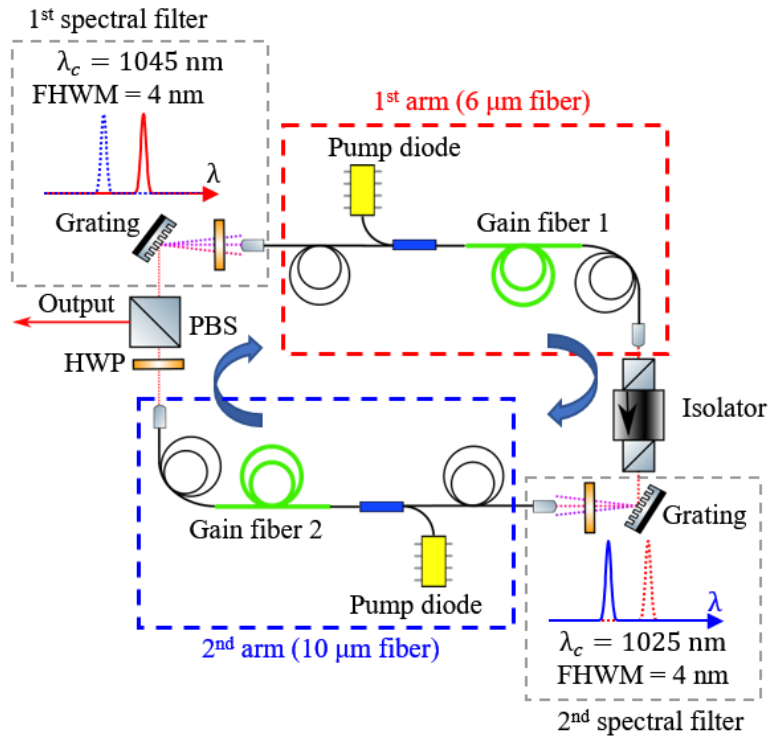

**Fig. 1** Schematic of the ring Mamyshev oscillator.

To initiate pulsation in the cavity, one needs to create an electric field fluctuation that is strong enough to sustain itself. To generate this fluctuation, we construct a "starting arm" (Fig. 2(a)) that forms an imbedded sub-cavity with only one filter. This bypassing of the second filter allows continuous-wave (CW) lasing inside the oscillator. A SA in the starting arm amplifies fluctuations to create the seed pulse. In our experimental realization, a non-PM fiber segment and polarization elements form an effective SA through NPE. When the starting arm is engaged (using a flip mirror; Fig. 2 (a)) and the SA is adjusted appropriately, noisy Q-switched pulses are generated in the auxiliary cavity. Fig. 2(b) shows a typical noisy Q-switched pulse train, measured at the auxiliary output in the starting arm, with the main cavity blocked. Figure 2(c) shows the spectrum averaged over 50 ms, i.e., over many pulses.

The noisy fluctuations from the starting arm initiate pulsation in the main cavity. Empirically, we find that if the spectrum of the noisy Q-switched state spans the two filter passbands, simply engaging and disengaging the flip mirror starts the mode-locking most reliably. The spectra of individual pulses may not span the filter passbands, so our empirical observation does not reflect a fundamental requirement for starting. The bandwidth of a noisy pulse is inversely proportional to the shortest temporal fluctuation, which will have high peak power on average. We therefore observe the most-reliable self-seeding with the broadest Q-switched spectra which can be obtained by properly setting the pump power in first arm and wave plates controlling SA of the starting arm. A more-detailed analysis of the starting mechanism will be the subject of future work. The main cavity consists only of PM fiber, making it environmentally-stable in steady-state operation (with the starting arm disengaged). While the use of NPE makes the starting arm sensitive to the environment, this does not affect the main cavity once the starting arm is disengaged. We anticipate that replacing the NPE with an environmentally-stable saturable absorber would serve comparably well, and allow even the starting arm to be PM. After initiation, mode-locking is self-sustaining and robust regardless of whether the starting arm is engaged.

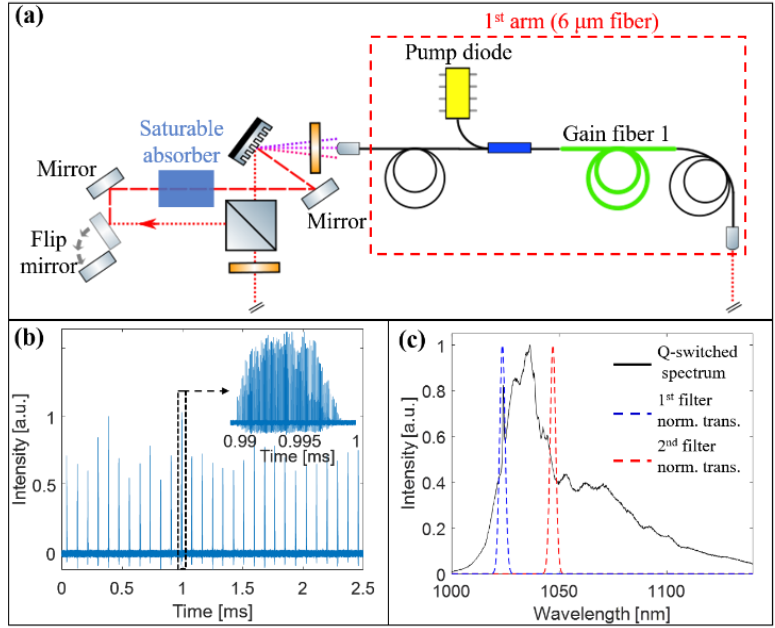

**Fig. 2.** (a) Schematic of the starting arm. The dotted red line indicates light path in the main cavity and red dashed line indicates light path when starting arm is engaged (only one arm of the main cavity is shown). (b) Noisy Q-switched pulse train. Inset: expanded view of the pulse marked by black dashed box. (c) The spectrum of the Q-switched pulse train (solid black curve). Dashed blue and red curves correspond to the normalized transmissions of the first and second filters respectively.

Once the oscillator is mode-locked, we adjust the output coupling ratio (i.e., the half-wave plate before the polarizing beam splitter in Fig. 1) to obtain the stable pulse train with highest pulse energy. The oscillator generates 4-ps chirped pulses with energy up to 190 nJ (Fig. 3(a)), with attempts to reach higher energies causing the pulse train to become unstable. Dechirping

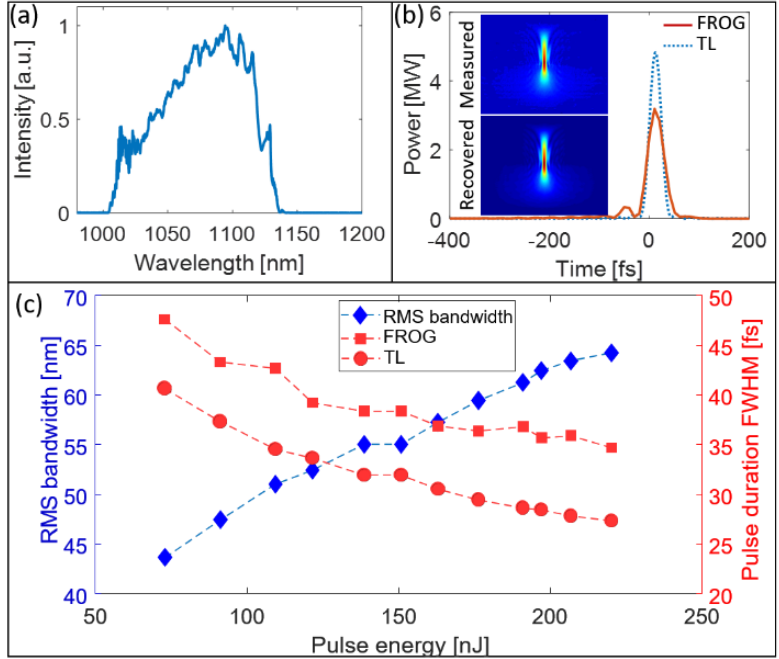

**Fig. 3** (a) Measured spectrum, (b) dechirped (red solid line) and TL (blue dotted line) pulses for 190 nJ pulse energy. Insets in (b) are measured and reconstructed FROG traces. (c) RMS bandwidth (blue diamonds), FWHM duration of the dechirped pulses (red squares) and TL duration (red circles) for pulse energies between 75 and 220 nJ.

the 190-nJ pulses with a grating compressor (1000 lines/mm) yields 35-fs pulses (Fig. 3(b)). The 75% efficiency of the compressor is accounted for in the peak power plotted in Fig. 3(b).

We systematically investigated the performance as a function of the pump power. The pulse energy is varied by changing the pump power in the 10-μm arm only. We verified single-pulse operation and the absence of CW breakthrough for all reported pulse energies. The bandwidth increases monotonically with pump power. Fig. 3(c) shows the variation of the RMS bandwidth (blue diamonds), the measured compressed pulse duration (red squares), and the calculated transform-limited (TL) pulse duration (red circles) with the pulse energy, as the pump power is increased. The increasing deviation between the measured and TL duration at high energy can be attributed to uncompensated cubic phase, with contributions from both the oscillator and the grating compressor. Calculations suggest that a compressor with third-order dispersion opposite that of an ordinary grating pair (e.g., a prism or grism [23] ) should allow the pulses to be compressed to their transform limit, which would improve the peak power to nearly 5 MW.

The peak power was verified by launching a known fraction of the dechirped pulse into 1 m of single mode fiber (SMF) and measuring the spectral broadening induced by self-phase modulation (SPM). The RMS bandwidth is compared with the results of numerical simulations that account for dispersion up to fourth order, SPM, self-steepening and Raman scattering (Fig. 4(a)). The simulated and measured bandwidths agree, confirming the peak power scale of Fig. 3(b).

The mode-locked state is quite stable on relevant time scales. The output pulse train was measured with a radio-frequency (RF) spectrum analyzer. The contrast between the fundamental frequency and secondary modulations offset about 17 kHz from the fundamental is 75 dB (Fig. 4(b)), which indicates stable mode-locking. After mode-locking is initiated with the auxiliary cavity, it routinely sustains itself for at least one day. Continuous operation without appreciable drift of the spectrum has been observed for up to three days, which is the longest period we have monitored.

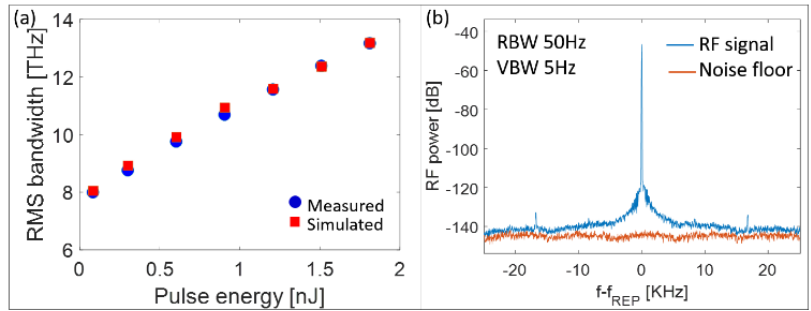

**Fig. 4** (a) Comparison of the measured spectral broadening of the dechirped pulse (blue circles) in 1 meter of SMF compared with simulations (red squares). (b) RF spectrum of the generated pulse train (blue curve) and instrument noise floor (red curve). RBW: resolution bandwidth; VBW: video bandwidth.

To understand the pulse evolution in the oscillator, we performed numerical simulations using the standard split-step method. The simulation includes the Kerr nonlinearity, Raman scattering, self-steepening and second- and third-order dispersion. For accurate modeling we found that it was critically important to model the Yb gain by solving the rate equations in the steady state [24]. The noisy seed pulse was generated by applying a pseudo-random spectral phase to a Gaussian spectrum with bandwidth of 20 nm, centered at 1030 nm. We verified that for given cavity parameters, the simulation converges to the same solution from different initial pulses. Fig. 5 shows results of the simulation for 190-nJ output pulses. The spectrum broadens dramatically in the gain fibers (Fig. 5(a,b)), as intended. The pulse duration also grows monotonically in both gain fibers. The filters abruptly truncate the pulse in both the spectral and temporal domains.

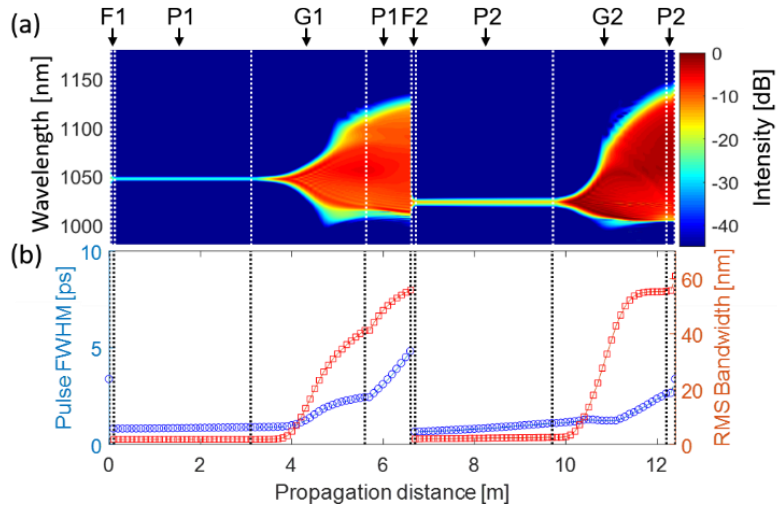

**Fig. 5** Numerical simulation results for 190 nJ output pulses. (a) Spectral evolution of the pulse in the cavity; P-passive fiber, G-gain fiber, F-filter (indices 1 and 2 correspond to the 6-μm and 10-μm arms, respectively). (b) Evolution of the pulse duration (blue circles) and RMS bandwidth (red squares).

For pulse energies up to 190 nJ, the simulations account well for the observed spectral shapes (Fig. 6(a)). At higher energies, the experimental spectrum develops structure, and this is accompanied by the instabilities in the pulse train mentioned above. The experimentally measured durations of the dechirped pulses are consistently longer than the simulated durations (respectively, blue squares, same data points as in Fig. 3 (c); and blue dashed curve in Fig. 6(b)). We attribute this discrepancy to the high-order dispersion coefficients of the actual fibers, which are not accounted for in simulations. The simulations predict stable pulse formation for energies up to 500 nJ, at which point pulse-splitting occurs. However, in experiments the pulse train becomes unstable above 190 nJ, and an abrupt loss of mode-locking occurs above 220 nJ. As this threshold is approached, the agreement between experiments and simulations breaks down, with a series of narrow peaks forming on the red side of the spectrum (Fig. 6(a)). Understanding this phenomenon may provide a route to the much higher pulse energies predicted numerically.

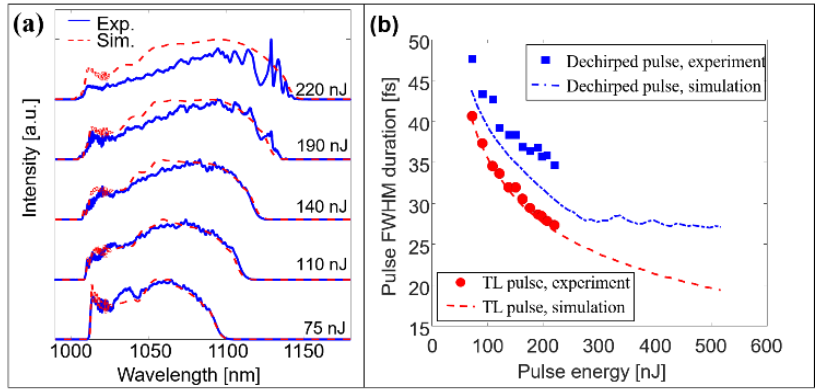

**Fig. 6** (a) Experimentally measured (blue) and simulated (red) spectra for range of pulse energies. (b) Measured and simulated pulse duration for dechirped and TL pulses.

The pulse energies and peak powers achieved here can be scaled further by established techniques such as the use of large-mode-area fibers [25,26] or divided-pulse amplification [27]. These may allow microjoule-level, 100-MW pulses to be obtained directly from an oscillator. It may also be interesting to combine the strong saturable absorber of the Mamyshev mechanism with the spatiotemporal evolution in multimode fibers [28].

In conclusion, we have demonstrated a fiber oscillator that generates 35-fs and 190-nJ pulses, for 3 MW peak power. Mode-locked operation is seeded by simply flipping a mirror to engage an auxiliary cavity, after which it is stable and robust. The self-seeding mechanism does not diminish the steady-state mode-locking performance. Numerical simulations suggest that the pulse energy can ultimately be increased up to ~500 nJ, and the pulse duration can be below 30 fs, resulting in ~20 MW peak power. The combination of a simple starting mechanism, environmental stability in the steady state, and the performance reported here make the Mamyshev oscillator an attractive source of ultrashort pulses for applications.

**References**
1. A. Chong, J. Buckley, W. Renninger, and F. Wise, "All-normal-dispersion femtosecond fiber laser," Opt. Express **14**, 10095 (2006).
2. W. H. Renninger and F. W. Wise, "Fundamental Limits to Mode-Locked Lasers: Toward Terawatt Peak Powers," IEEE J. Sel. Top. Quantum Electron. **21**, 63–70 (2015).
3. J. Jeon, J. Lee, and J. H. Lee, "Numerical study on the minimum modulation depth of a saturable absorber for stable fiber laser mode locking," J. Opt. Soc. Am. B **32**, 31 (2015).
4. C. Ma, X. Tian, B. Gao, and G. Wu, "Numerical simulations on influence of the saturable absorber in Er-doped fiber laser," Opt. Commun. **410**, 941–946 (2018).
5. U. Keller, K. J. Weingarten, F. X. Kartner, D. Kopf, B. Braun, I. D. Jung, R. Fluck, C.

Honninger, N. Matuschek, and J. Aus der Au, "Semiconductor saturable absorber mirrors (SESAM's) for femtosecond to nanosecond pulse generation in solid-state lasers," IEEE J. Sel. Top. Quantum Electron. **2**, 435–453 (1996).
6. S. Y. Set, H. Yaguchi, Y. Tanaka, and M. Jablonski, "Ultrafast Fiber Pulsed Lasers Incorporating Carbon Nanotubes," IEEE J. Sel. Top. Quantum Electron. **10**, 137–146 (2004).
7. A. Martinez and Z. Sun, "Nanotube and graphene saturable absorbers for fibre lasers," Nat. Photonics **7**, 842–845 (2013).
8. R. H. Stolen, J. Botineau, and A. Ashkin, "Intensity discrimination of optical pulses with birefringent fibers," Opt. Lett. **7**, 512 (1982).
9. N. J. Doran and D. Wood, "Nonlinear-optical loop mirror," Opt. Lett. **13**, 56 (1988).
10. M. E. Fermann, F. Haberl, M. Hofer, and H. Hochreiter, "Nonlinear amplifying loop mirror," Opt. Lett. **15**, 752 (1990).
11. C. J. Saraceno, C. Schriber, M. Mangold, M. Hoffmann, O. H. Heckl, C. R. Baer, M. Golling, T. Südmeyer, and U. Keller, "SESAMs for High-Power Oscillators: Design Guidelines and Damage Thresholds," IEEE J. Sel. Top. Quantum Electron. **18**, 29–41 (2012).
12. K. Viskontas, K. Regelskis, and N. Rusteika, "Slow and fast optical degradation of the SESAM for fiber laser mode-locking at 1 μm," Lith. J. Phys. **54**, (2014).
13. P. V. Mamyshev, "All-optical data regeneration based on self-phase modulation effect," in *24th European Conference on Optical Communication. ECOC '98 (IEEE Cat. No.98TH8398)* (Telefonica, n.d.), Vol. 1, pp. 475–476.
14. U. Keller, T. H. Chiu, and J. F. Ferguson, "Self-starting femtosecond mode-locked Nd:glass laser that uses intracavity saturable absorbers," Opt. Lett. **18**, 1077 (1993).
15. M. Piché, "Mode locking through nonlinear frequency broadening and spectral filtering," in *Proc. SPIE*, M. Piche and P. W. Pace, eds. (1994), pp. 358–365.
16. S. Pitois, C. Finot, L. Provost, and D. J. Richardson, "Generation of localized pulses from incoherent wave in optical fiber lines made of concatenated Mamyshev regenerators," J. Opt. Soc. Am. B **25**, 1537 (2008).
17. K. Sun, M. Rochette, and L. R. Chen, "Output characterization of a self-pulsating and aperiodic optical fiber source based on cascaded regeneration," Opt. Express **17**, 10419 (2009).
18. T. North and M. Rochette, "Regenerative self-pulsating sources of large bandwidths," Opt. Lett. **39**, 174 (2014).
19. K. Regelskis, J. Želudevičius, K. Viskontas, and G. Račiukaitis, "Ytterbium-doped fiber ultrashort pulse generator based on self-phase modulation and alternating spectral filtering," Opt. Lett. **40**, 5255 (2015).
20. I. Samartsev, A. Bordenyuk, and V. Gapontsev, "Environmentally stable seed source for high power ultrafast laser," in A. L. Glebov and P. O. Leisher, eds. (2017), p. 100850S.
21. Z. Liu, Z. M. Ziegler, L. G. Wright, and F. W. Wise, "Megawatt peak power from a Mamyshev oscillator," Optica **4**, 649 (2017).
22. S. Pitois, C. Finot, and L. Provost, "Asymptotic properties of incoherent waves propagating in an all-optical regenerators line," Opt. Lett. **32**, 3263 (2007).
23. S. Kane and J. Squier, "Grism-pair stretcher–compressor system for simultaneous second- and third-order dispersion compensation in chirped-pulse amplification," J. Opt. Soc. Am. B **14**, 661 (1997).
24. R. Paschotta, J. Nilsson, A. C. Tropper, and D. C. Hanna, "Ytterbium-doped fiber amplifiers," IEEE J. Quantum Electron. **33**, 1049–1056 (1997).
25. M. Baumgartl, B. Ortaç, C. Lecaplain, A. Hideur, J. Limpert, and A. Tünnermann, "Sub-80 fs dissipative soliton large-mode-area fiber laser," Opt. Lett. **35**, 2311 (2010).
26. M. Baumgartl, C. Lecaplain, A. Hideur, J. Limpert, and A. Tünnermann, "66 W average

power from a microjoule-class sub-100 fs fiber oscillator," Opt. Lett. **37**, 1640 (2012).
27. E. S. Lamb, L. G. Wright, and F. W. Wise, "Divided-pulse lasers," Opt. Lett. **39**, 2775 (2014).
28. L. G. Wright, D. N. Christodoulides, and F. W. Wise, "Spatiotemporal mode-locking in multimode fiber lasers," Science (80-. ). **358**, 94–97 (2017).